\documentclass[preprint,aps,showpacs,amsmath]{revtex4}
\usepackage{graphicx}
\usepackage{float}
\newcommand{\y }{\'{\i}}
\newcommand{\be }{\begin{equation}}
\newcommand{\ee }{\end{equation}}

\begin{document}

\title{Decay of metastable phases in a model for the catalytic oxidation of CO}
\author{Erik Machado}
\author{Gloria M.~Buend\y a}
\affiliation{Physics Department, Universidad Sim\'on Bol\y var,\\
Apartado 89000, Caracas 1080, Venezuela}
\author{Per Arne Rikvold}
\affiliation{Center for Materials Research and Technology, 
School of Computational Science, National High Magnetic Field
Laboratory, and Department of Physics,
Florida State University, Tallahassee, FL 32306-4350, USA}

\date{\today}

\begin{abstract}
We study by kinetic Monte Carlo simulations the dynamic behavior of a
Ziff-Gulari-Barshad model with CO desorption for the
reaction CO + O $\rightarrow$ CO$_2$ on a catalytic surface.
Finite-size scaling analysis of the fluctuations and 
the fourth-order order-parameter cumulant show that below a 
critical CO desorption rate, the model exhibits a nonequilibrium first-order 
phase transition between low and high CO coverage phases. We calculate
several points on the coexistence curve.
We also measure the metastable lifetimes associated
with the transition from the low CO coverage phase to
the high CO coverage phase, and {\it vice versa\/}. 
Our results indicate that the transition process   
follows a mechanism very similar to the decay of metastable phases associated
with {\it equilibrium\/} first-order phase transitions and can be described 
by the classic Kolmogorov-Johnson-Mehl-Avrami 
theory of phase transformation by nucleation and growth. In the present case, 
the desorption parameter plays the role of temperature, and the
distance to the coexistence curve plays the role of an external field or 
supersaturation.  We identify two distinct regimes, 
depending on whether the system is far from or close to the coexistence curve, 
in which the statistical properties and the
system-size dependence of the lifetimes are different, corresponding 
to multidroplet or single-droplet decay, respectively. 
The crossover between the two regimes 
approaches the coexistence curve logarithmically
with system size, analogous to the behavior of the crossover between
multidroplet and single-droplet metastable decay near an equilibrium 
first-order phase transition.

\end{abstract}


\maketitle

\section{Introduction}
\label{sec:I}

The study of phase transitions and critical phenomena in
nonequilibrium systems is a subject of great interest. In
particular, the study of surface reaction models has attracted
considerable attention \cite{general}. These models not only
exhibit rich and complex behavior, but they can also explain a
wide range of experimental results associated with catalysis and
could be very useful for designing more efficient processes. The
potential applications of improved catalytic reactions are a
powerful reason to pursue this line of research \cite{catalysis}.
Unlike the decay of metastable phases near first-order equilibrium 
phase transitions, the decay of metastable  
phases near nonequilibrium phase transitions still 
lacks a well-established theoretical framework. 

The Ziff, Gulari, and Barshad (ZGB) model is a lattice-gas adsorption-reaction
model that describes some kinetic aspects of the catalytic oxidation
of carbon monoxide on a crystal surface \cite{ziff86}. The ZGB
model assumes that the reaction between CO and O$_2$ on a
surface proceeds according to the Langmuir-Hinshelwood process:
\begin{eqnarray}
\text{CO(g)} + \text{S} & \rightarrow & \text{CO(a)}
\nonumber \\
\text{O}_2 + 2\text{S} & \rightarrow & 2\text{O(a)}
\nonumber \\
\text{CO(a)} + \text{O(a)} & \rightarrow & \text{CO}_2\text{(g)} + 2\text{S}
\;,
\nonumber
\end{eqnarray}
where S is an empty site on the surface, and (g) and (a) refer to the gas
and adsorbed phase, respectively.
The process is controlled by a single parameter $y$, which represents the
probability that the next molecule arriving at the surface is 
CO, i.e., it is proportional to the partial pressure of CO. The
 model exhibits two kinetic phase transitions, a continuous one at $y=y_1$
and a discontinuous one at $y=y_2$, where $y_1 < y_2$. 
When $y<y_1$, all the sites
become occupied by oxygen, the so-called oxygen-poisoned state. If
$y>y_2$, all the sites become occupied by CO molecules, 
the so-called CO-poisoned state. Real
systems do not possess an oxygen-poisoned state because oxygen does
not impede the adsorption of CO. However, transitions between
states of low and high CO coverage $\theta_\text{CO}$ (where 
$\theta_\text{CO}$ is the fraction of surface sites
occupied by CO) have been observed experimentally \cite{ehsasi89,X1}. At low
temperatures, as $y$ increases, there is a discontinuous increase in 
$\theta_{\rm CO}$, accompanied by a discontinuous drop in
the CO$_2$ production rate. Above a critical temperature
the discontinuities disappear, and the  CO$_2$ production decreases
continuously. This type of behavior can be reproduced by modifying
the ZGB model to include a CO desorption rate, $k$ 
\cite{fisher89,kaukonen89,albano92,brosilow92},
a model we for brevity will call the ZGB-k model \cite{machado04}. 
For this model, there is a
distinction between high and low CO-coverage phases only for $k$
below a critical value $k_c$, while above $k_c$ the CO coverage
varies smoothly with $y$. Thus, the transition value $y_2$ becomes a
function of $k$, corresponding to a coexistence curve $y_{2}(k)$ that
terminates at the critical point $y_2(k_c)$ \cite{tome93, ziffb92}. The
ZGB-k model does not have a totally poisoned CO state, and hysteresis
is observed in $\theta_\text{CO}$ as $y$ is varied close to
$y_2(k)$ \cite{tome93, machado04}. 
This hysteresis is associated with well-defined metastable phases of
the model. 

The main aim of this paper is to understand the dynamical response of the
model near the discontinuous transition. We present an extensive finite-size 
scaling analysis of results from kinetic Monte Carlo simulations 
that indicates, more conclusively than previous
studies, that the system undergoes a first-order nonequilibrium phase 
transition along
a coexistence curve that terminates at a critical point. We calculate 
several points on the  coexistence curve. Next we measure 
the lifetimes of the metastable phases associated with the decay
from the low (high) CO coverage phase to the high (low) coverage phase. We
find that the statistics of the metastable lifetimes are well described 
by the Kolmogorov-Johnson-Mehl-Avrami (KJMA) \cite{KOLM37,JOHN39,AVRAMI}
theory of phase transformation by nucleation and growth near a
first-order {\it equilibrium\/} phase transition.

The outline of the paper is as follows. In Sec.~\ref{sec:ModS} we define 
the model and
describe the Monte Carlo simulation techniques used. 
In Sec.~\ref{sec:R} we present
and discuss the numerical results obtained: in Sec.~\ref{sec:CC} we show how we
calculate the coexistence curve and present a finite-size scaling analysis
of the fluctuations and of the fourth-order cumulant of the order parameter;
in Sec.~\ref{sec:MS} we present 
the measurements of the lifetimes of the metastable
states associated with the transition and show how their behavior is
described by the KJMA theory.
Our conclusions are summarized in Sec.~\ref{sec:conc}.

\section{Model and Simulation}
\label{sec:ModS}

\begin{figure}
\centering\includegraphics[scale=.5, clip]{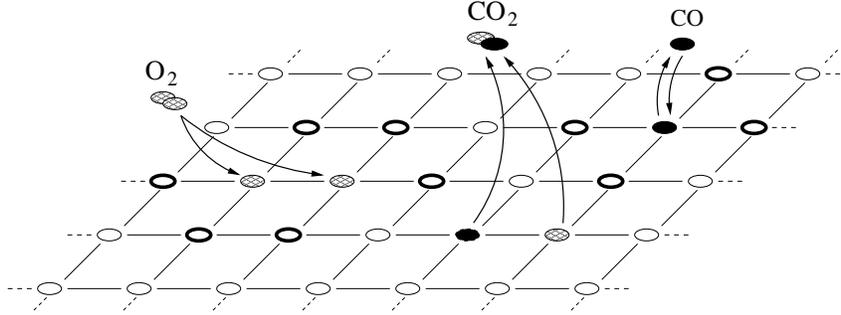}
\caption[]{Schematic representation of the algorithm.
See discussion in the text.}
\label{model}
\end{figure}

The ZGB model with desorption is simulated on a square lattice of
linear size $L$ that represents the catalytic surface. A Monte Carlo
simulation generates a sequence of trials: CO or O$_2$ adsorption
with probability $1-k$ and CO desorption with probability $k$. In
the case of adsorption  a CO or O$_2$ molecule is selected
with probability $y$ and $1-y$ respectively \cite{ziff86,tome93}.
These probabilities are the relative impingement rates of
both molecules and are proportional to their partial pressures.
The algorithm works in the following way. A site $i$ is selected
at random. In the case of desorption, if $i$ is occupied by CO the site
is vacated and the trial ends, if not the trial also ends.
In the case of adsorption, if a CO
molecule is selected it can be adsorbed at the empty site $i$ if
none of its nearest neighbors are occupied by an O atom.
Otherwise, one of the occupied O neighbors is selected at random and
removed from the surface, leaving $i$ and the selected neighbor
vacant. This move simulates the CO + O $\rightarrow$ CO$_2$ surface
reaction following the adsorption of CO.
O$_2$ molecules can be adsorbed only if a pair of nearest-neighbor
sites are vacant. If the adsorbed molecule is selected to be 
O$_2$ a nearest neighbor of $i$, $j$, is selected at random, and if it
is occupied the trial ends. If both $i$ and $j$ are empty, the
trial proceeds, and the O$_2$ molecule is adsorbed and dissociates
into two O atoms. If none of the remaining neighbors of $i$ is
occupied by a CO molecule, one O atom is located at $i$, and if
none of the neighbors of $j$ is occupied by a CO molecule, then the
other O is located at $j$. If any neighbors of $i$ are occupied by
a CO, then one is selected at random to react with the O at $i$ such
that both sites are vacated. The same
 reaction happens at site $j$ if any of its neighbors are filled
with a CO molecule. This process mimics the CO + O $\rightarrow$
CO$_2$ surface reaction following O$_2$ adsorption.
A schematic representation of this
algorithm is shown in Fig.~\ref{model}.
We emphasize that the ZGB model, both with and without CO desorption,
is an intrinsically nonequilibrium model 
that is fully defined by these dynamic rules. 
In contrast to systems considered in
equilibrium thermodynamics, its properties are {\it not\/}
derived from a Hamiltonian. We shall return to this point in Sec.~\ref{sec:MS}. 

For our simulations we assume periodic boundary conditions. The
time unit is one Monte Carlo Step per Site (MCSS), in which each
site, on average, is visited once. 
For measurements of stationary quantities, the system was allowed to
reach stationarity before data were recorded for analysis. 
Averages were taken over $10^3$ independent simulation runs.

\section{Results}
\label{sec:R}

\begin{figure}
\centering\includegraphics[scale=.5, clip]{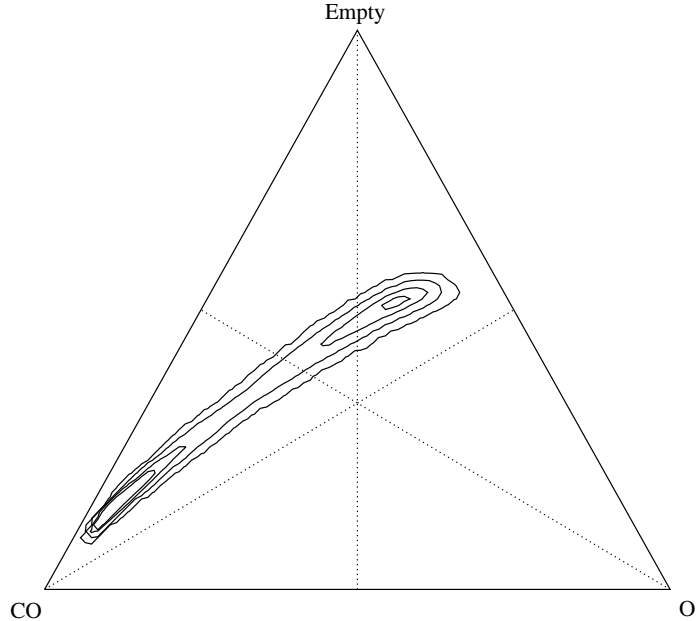}
\caption[]{Contour plot of the projection of $10^6$ MCSS of simulation
onto the ternary phase diagram for $k=0.02$, $y=0.5332$, and $L=100$.}
\label{triplot}
\end{figure}

We use a standard ternary phase diagram to plot the
fraction of sites occupied by CO molecules: the CO coverage,
$\theta_\text{CO}$; the O coverage, $\theta_\text{O}$, and the fraction of
empty sites, $\theta_\text{E}$. In Fig.~\ref{triplot} we
present a contour plot of a histogram based on 
the projection of 
$10^6$ MCSS onto the plane of the phase diagram.
For the chosen parameters and observation times the system
undergoes several transitions between the low and high CO coverage
phases. From the fact that the set of phase points is nearly
parallel to the line $\theta_\text{O}/\theta_\text{E}=1/2$, it is evident that
the CO coverage gives more information about the kinetic phase
transitions than $\theta_\text{O}$ or $\theta_\text{E}$.

\subsection{Determination of the Coexistence Curve}
\label{sec:CC}

We estimate $P(\theta_\text{CO})$, the probability distribution for
$\theta_{\rm CO}$, by recording the number of times $N_i$ that the
coverage fell in the intervals $[0, \epsilon), [\epsilon, 2\epsilon),
...,[1-\epsilon, 1]$ ($\epsilon = 0.01$), such that $\sum_i N_i = N$
is the total number of MCSS.
\begin{figure}
\input{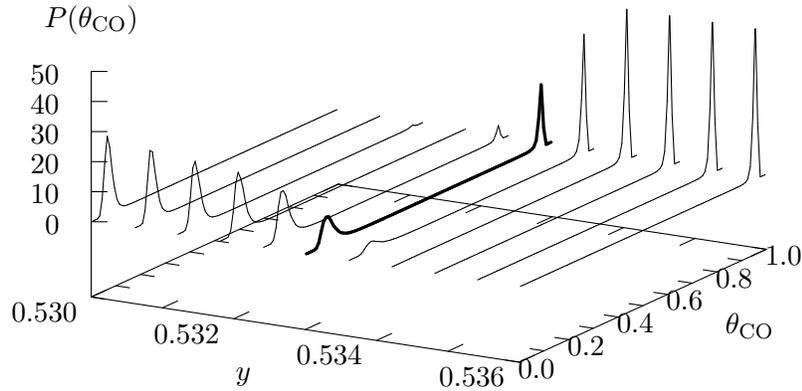}
\caption[]{Order-parameter probability distribution, 
$P(\theta_\text{CO})$, shown vs $y$ for $k=0.03$ and $L=100$.
The distribution for the value of $y$ closest to the coexistence value 
is shown with a bold line.}
\label{histos3}
\end{figure}
Then, the probability that $\theta_{\rm CO}$ has a value
in the interval $[(i-1)\epsilon, i\epsilon)$ is
$P_i={N_i}/{N\epsilon}$, 
such that
\begin{equation}
\int_0^1 P(\theta_\text{CO}) d\theta_\text{CO}=\sum_i P_i\epsilon=1.
\end{equation}
In Fig.~\ref{histos3} we show $P(\theta_\text{CO})$
versus $y$. In the regions
(below and above $y_2$) where the histograms are unimodal,
the system consists of one single phase. For a very narrow range of $y$,
the histograms are bimodal, indicating two distinct phases.
At the coexistence point $y_2(k)$, the areas under both peaks
are equal \cite{borg90,land00}.

We define a measure of the
fluctuations in $\theta_\text{CO}$ in a $L \times L$ system in the standard way as
\begin{equation}
X_L=L^2(\langle \theta_\text{CO}^2\rangle_{L} 
-\langle \theta_\text{CO}\rangle^{2}_{L})
\;,
\end{equation}
where
\begin{equation}
\langle \theta_\text{CO}^n\rangle_{L}=
\int_0^1 \theta_\text{CO}^n P(\theta_\text{CO}) d\theta_\text{CO}
\;.
\end{equation}
We measure $X_L$ as a function of $y$ for a
fixed value of $k$ and several values of $L$. At
a first-order equilibrium phase
transition, the order-parameter fluctuations increase with
the system size, such that the  maximum value of $X_L \sim L^d$, 
where $d$ is the spatial dimension of the system 
\cite{bind84,land00,challa86,nien75}.
We will take the same scaling behavior to indicate a 
{\it nonequilibrium\/} first-order transition. 

\begin{figure}
\centering\includegraphics[scale=.5, clip]{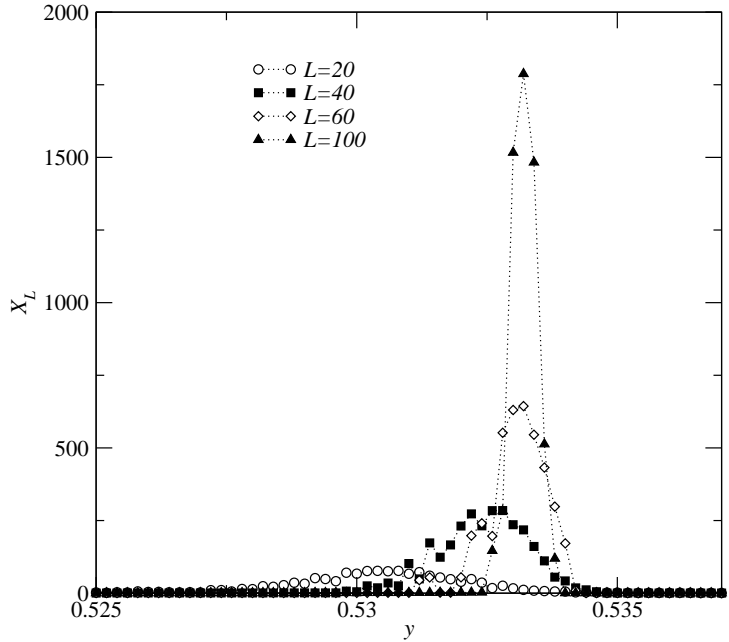}
\caption[]{The order-parameter fluctuation measure
$X_L$, shown vs $y$ for $k=0.02$ and four system
sizes. The dotted lines are guides to the eye.
The values of $X_L$ have an error of approximately 5\%.}
\label{fluct2}
\end{figure}

In Fig.~\ref{fluct2} we show $X_L$ vs $y$ for four system
sizes at $k=0.02$. For the four
values of $L$ used, $X_L$ displays a clear peak, which moves and increases in
height with increasing $L$.

\begin{figure}
\centering\includegraphics[scale=.5, clip]{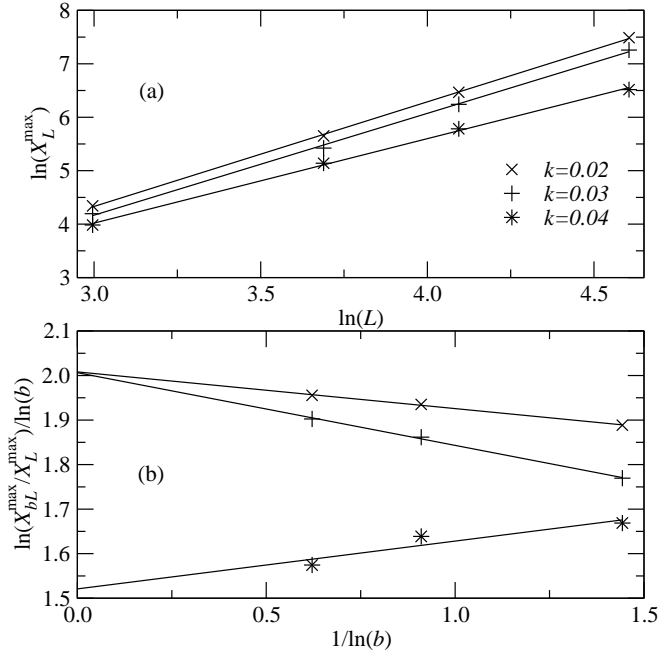}
\caption[]{(a) Plot of $\ln(X_L^\text{max})$ vs $\ln(L)$ and
(b) plot of $\ln(X_{bL}^\text{max}/X_L^\text{max})/\ln(b)$ with $L=20$ vs $1/\ln(b)$,
both for several values of $k$ and including all four system sizes.
$X_L^\text{max}$ is the maximum value of $X_L$, taken from figures
similar to Fig.~\ref{fluct2}. The straight lines are
the best linear fits to the data and give $X_L^\text{max}\sim L^{d^{\prime}}$
with (a) $d^{\prime}=1.96\pm 0.02, 1.91\pm 0.05$, and $1.58\pm 0.04$; and
(b) $d^{\prime}=2.008\pm 0.005, 2.006\pm 0.008$, and $1.52\pm 0.04$; for
$k=0.02, 0.03$, and 0.04, respectively.}
\label{FSS_fluct_max}
\end{figure}

In Fig.~\ref{FSS_fluct_max}(a) we plot $\ln(X_L^\text{max})$
versus $\ln(L)$ for several values of $k$.
A linear fit indicates a power-law divergence with $L$,
such that  the maximum value scales as
$X_L^\text{max}\sim L^{d^{\prime}}$ with  
$d^{\prime}=1.96\pm 0.02, 1.91\pm 0.05$, and $1.58\pm 0.04$ for
$k=0.02, 0.03$, and 0.04, respectively.
A different method to extract the power-law exponent, which has some advantage
in eliminating the effects of a nonsingular background term
(as in $X_L = f + gL^{d^{\prime}}$ with $f$ and $g$ constants), is to consider
\begin{equation}
\ln\left[\frac{X_{bL}^\text{max}}{X_L^\text{max}}\right]/\ln b=
d^{\prime} + {\cal O}(1/\ln b)
\label{eq:bLX}
\end{equation}
with $L$ fixed at a relatively small value (here, $L=20$), and $b>1$.
For large $L$ and $b$,
the correction term is proportional to $f/(g \ln b)$, so that the exponent
can be estimated by plotting the left-hand-side of Eq.~(\ref{eq:bLX}) vs
$1/\ln b$ and extrapolating to $1/\ln b = 0$, as in
Fig.~\ref{FSS_fluct_max}(b). The resulting estimates are
$d^{\prime}=2.008\pm 0.005, 2.006\pm 0.008$, and $1.52\pm 0.04$ for
$k=0.02, 0.03$, and 0.04, respectively. 

These results indicate that the system undergoes a first-order phase transition,
$d^{\prime}\approx d=2$, at $y= y_2(k)$, 
generating a coexistence curve that terminates
at a critical point $0.03< k_{c}<0.04$. A previous estimate based on a study of
the fractal dimension of the interface between the two phases
gives $0.039 < k_{c} < 0.04$ \cite{brosilow92}.
Another study, which estimates $k_{c}(L)$ as the value of $k$ 
where the double-peaked histograms become single-peaked,
gives $k_{c}(L \rightarrow \infty)=0.04060$  \cite{tome93}. However, 
from the known two-peaked shape of the order-parameter distribution at the 
equilibrium Ising critical point \cite{bruce85}, we believe 
that this method should 
yield a slight overestimate of $k_c$. Reference \cite{tome93} also 
reports preliminary results on the fourth-order cumulant of the CO coverage 
that are consistent with an Ising-like critical point at $k_{c}=0.0406$. 
However, the cumulants
were calculated only for very small lattice sizes ($L=10$, 20, and $40$) and 
do not constitute a definite
proof of the location of the critical point, 
as the authors duly point out \cite{tome93}.
Below we present a similar cumulant study, but with larger lattice sizes.

We also calculated the relation 
between the value of the probability distribution
$P(\theta_\text {CO})$ at either 
of the peaks of the bimodal distribution, $P_\text{max}$,
and its value at the local minimum between the peaks, $P_\text{min}$. 
For a first-order 
equilibrium phase transition these quantities satisfy the relation 
\begin{equation}
\frac{P_\text{max}}{P_\text{min}}\propto \exp(cL) \;,
\label{eq:Pratio}
\end{equation}
where $c$ is proportional to the equilibrium interface tension between
the two phases. If the relation is applied to the present system, we
would expect $c$ to be positive and decrease with increasing $k$. Our results
indicate that $c(k)$ is positive only for $k<0.03$, 
suggesting that $0.02<k_{c}< 0.03$.
These results corroborate again that the system has a 
first-order phase transition
for small $k$. However, they suggest a much lower value for $k_c$ 
than indicated by our other techniques and the previous results by others 
\cite{brosilow92,tome93}.
We find this result quite interesting and believe it may be due to
several reasons. Most obvious is the significant difficulty in locating
and measuring the peaks, which are extremely narrow in $y$. Perhaps more
significant is the fact that this non-Hamiltonian nonequilibrium system
does not possess a well-defined surface tension that could be associated
with the parameter $c$ in Eq.~(\ref{eq:Pratio}). This point will be
further discussed in Sec.~\ref{sec:MS}, where we also show that the
cluster interfaces in this system are much rougher than in conventional
Hamiltonian systems. As a result, we do not consider the method for 
determining $k_c$, based on Eq.~(\ref{eq:Pratio}), very accurate.  
The fact that different techniques give different 
results are an indication of the difficulties associated
with locating $k_c$ in this model, even for relatively large systems.

\begin{figure}
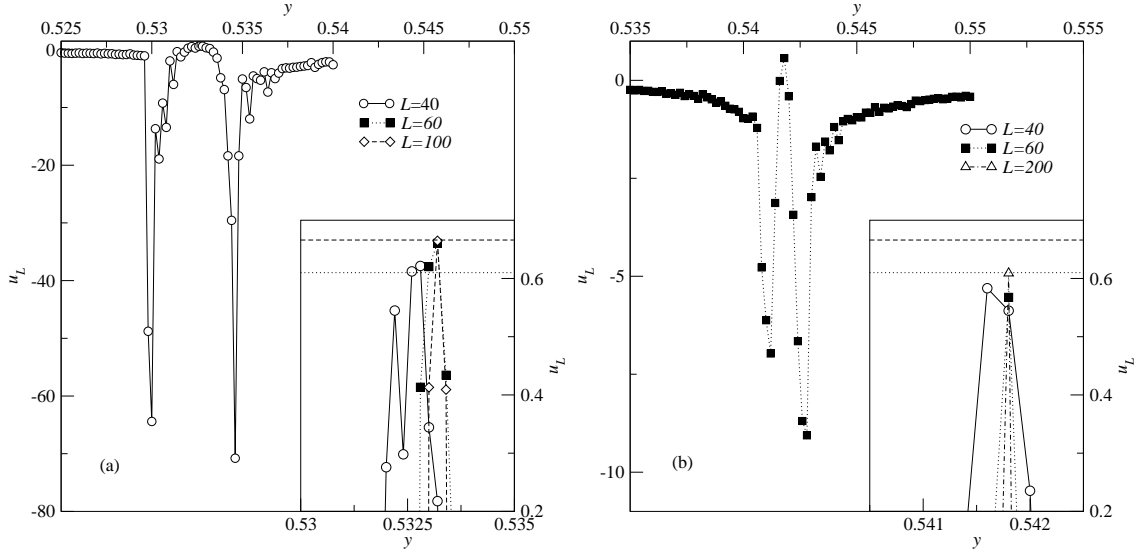

\centering\includegraphics[scale=.4, clip]{fig6a.eps}
\centering\includegraphics[scale=.4, clip]{fig6b.eps}
\caption[]{Dependence of the fourth-order reduced cumulant $u_L$ on $y$,
for (a) $k=0.02$ and (b) $k=0.04$. The $P(\theta_\text{CO})$ histograms 
indicate that the minima of $u_L$ correspond to the transitions from 
unimodal to bimodal distribution.
The maximum between them gives the coexistence point $y_2(k,L)$.
The horizontal lines in the insets correspond to $u_L=2/3$ (dashed) and
$u_\text{Ising}^*\approx 0.61$ (dotted).
}
\label{uL_vs_y}
\end{figure}

A useful tool for detecting phase transitions in simulations of finite
{\it equilibrium\/} systems is the fourth-order reduced cumulant of the
order parameter \cite{land00,bind84,challa86}.
For $\theta_{\rm CO}$ it takes the form
\begin{equation}
u_L = 1 - \frac{\mu_4}{3\mu_2^2},
\end{equation}
where,
\begin{equation}
\mu_n =  \langle (\theta_\text{CO}-\langle
\theta_\text{CO}\rangle )^n\rangle=
\int_0^1 (\theta_\text{CO}-\langle \theta_\text{CO}\rangle)^n
P(\theta_\text{CO}) d\theta_\text{CO},
\end{equation}
is the $n$th central moment of the CO coverage.
The equal-area bimodal distribution corresponding to coexistence
yields a positive maximum for the cumulant vs $y$,
flanked on either side by negative minima and approaching zero far away from
the transition.
Since the cumulant essentially is a tool to determine the shape of the
order-parameter distribution, it can also be used for nonequilibrium
phase transitions, such as the one studied here.
The maxima of $u_L$ define the
$L$-dependent coexistence line, $y_2(k,L)$. In Fig.~\ref{uL_vs_y}
we show the dependence of $u_L$ on $y$ for several
values of $L$ and for $k=0.02$ and $k=0.04$, respectively. When $k=0.02$,
the maximum value of $u_L$ is very close to $2/3$, however when $k=0.04$
the maximum is very close to $0.61$, consistent with 
the proximity of an Ising-like critical point.

\begin{figure}
\centering\includegraphics[scale=.5, clip]{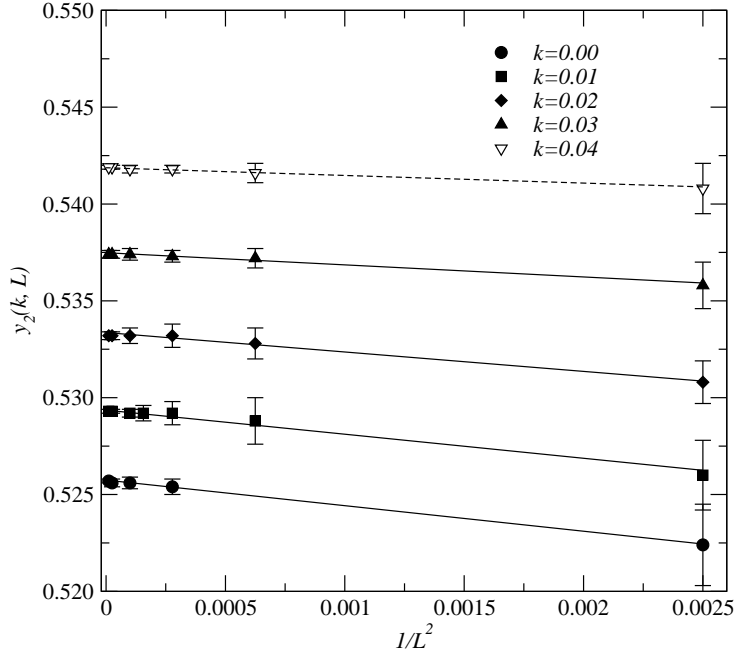}
\caption[]{Critical CO adsorption rate $y_2(k,L)$, shown vs $1/L^2$
for several values of the desorption rate $k$.
The straight lines are weighted least-squares fits, yielding the estimated
$y_2(k)$ for $L=\infty$.}
\label{size_scaling}
\end{figure}

The finite-size scaling theory of \textit{equilibrium}
first-order phase transitions implies that the shift in the
position of the transition in a finite system of linear size $L$ 
with periodic boundary conditions is inversely
proportional to the system volume, $L^d$ 
\cite{challa86,nien75,FISH82}
(here the dimension $d=2$).
Although there is no analogous scaling theory for the present nonequilibrium
system, we here attempt to use the same scaling relation,
\begin{equation}
y_2(k,L)-y_2(k) \propto L^{-2},
\end{equation}
where $y_2(k)$ is the transition value of the  CO adsorption rate $y$ in the
infinite-$L$ limit.
In Fig.~\ref{size_scaling} we plot $y_2(L)$
vs $1/L^2$ for $L=20, 40, 60, 80, 100, 200$, and $300$ for several
values of $k$. The error bars are calculated as the half width of
the peaks of the $u_L$ vs $y$ curves \cite{bind84,challa86}.
As seen from the figure, the
points fall very close to the straight line representing a weighted
least-squares fit, yielding a 
good estimate (within $10^{-2}\,$\% ) of $y_2(k)$ for each
$k$ when $L\rightarrow \infty$.
Our result $y_2=0.5257(3)$ for $k=0$ is consistent with previous
studies that give $y_2=0.52560(1)$ \cite{ziffb92} and
$y_2=0.52583(9)$ \cite{Monetti01} for $L \rightarrow \infty$.

\begin{figure}
\centering\includegraphics[scale=.5, clip]{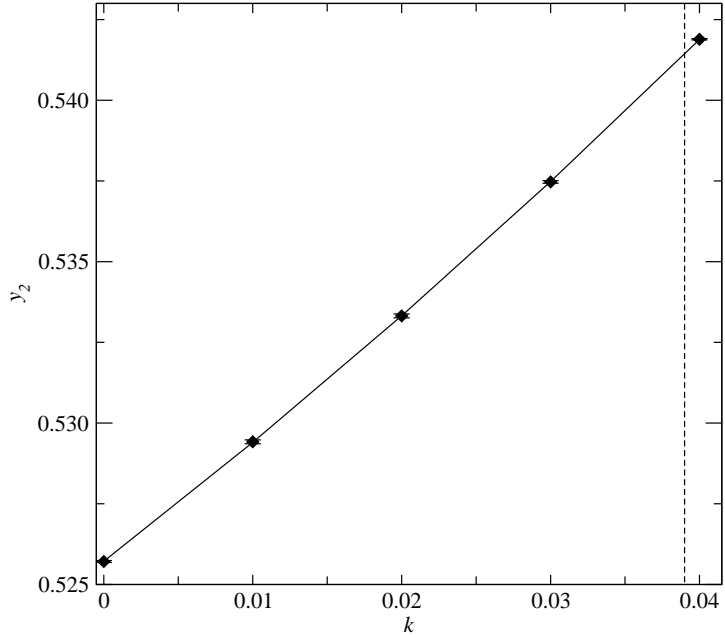}
\caption[]{Some points of the coexistence curve, analogous to the pressure vs 
temperature phase diagram for a fluid in equilibrium. 
The continuous line represents
a quadratic fit to the data. The vertical dashed line indicates the estimate 
$k_c \approx 0.039$ from Ref.~\protect\cite{brosilow92}.}
\label{coex}
\end{figure}

In Fig.~\ref{coex} we show several points on the coexistence curve
between the low and high CO 
coverage phases. The coexistence curve is almost linear with
only a slight curvature.
By extrapolating a quadratic fit of the data
to $k=0.04060$ we obtain $y_2=0.542(3)$ in agreement with the
previous result $y_2=0.54212(10)$ for $k_c$ \cite{tome93}.

\subsection{Metastability}
\label{sec:MS}
 In this section we calculate the time
associated with the decay from the low (high) CO coverage phase to
the high (low) CO coverage phase, $\tau_\text{p}$
($\tau_\text{d}$), and 
determine its dependence on the CO pressure and the system size.
In order to do this, we prepare the system with initial pressure
$y=y_\text{w}$ such that $y_{1} < y_\text{w} < y_{2}(k)$
($y_\text{w} > y_{2}(k)$) and then suddenly change $y$, such that $y >
y_{2}(k)$ ($y < y_{2}(k)$). Then the initial low (high)- coverage
phase becomes metastable and eventually decays to the high
(low)-coverage phase. The system is considered to have left the
metastable phase once its coverage reaches a certain cutoff value
$\theta_\text{CO}^*$. To avoid that recrossing events back to the
metastable phase are mistaken for decay events, the cut-off is
selected such that it is not too close to the metastable coverage
value. The statistics of the decay times
are analyzed for $n=500$ independent runs.

\begin{figure}
\centering\includegraphics[scale=.5, clip]{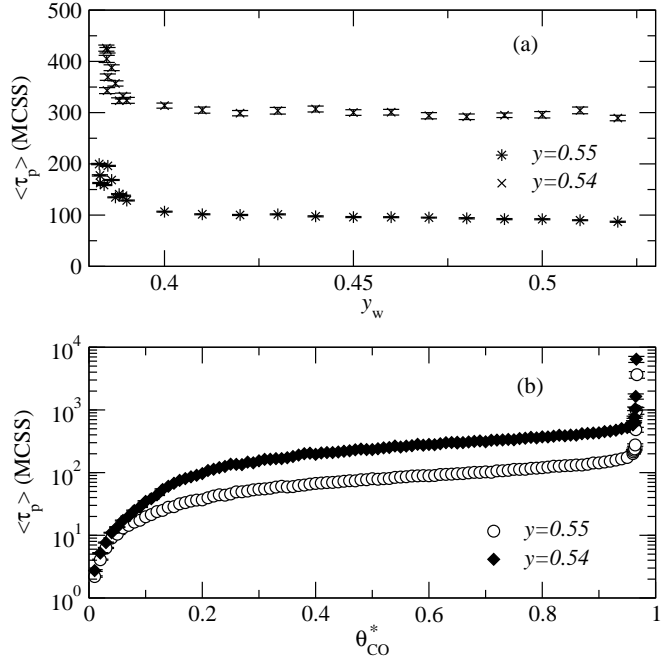}
\caption[]{$\langle\tau_\text{p}\rangle$ as a function of (a) $y_\text{w}$ with
$\theta_\text{CO}^*=0.65$, and (b) $\theta_\text{CO}^*$
with $y_\text{w}=0.475$; for $k=0.02$, $L=100$ and two different values of $y$.}
\label{w_c_p}
\end{figure}

Since the value of $y_\text{w}$ that determines how far the
initial system is from the transition point $y_2(k)$ is somewhat
arbitrary, it is necessary to evaluate how the decay times depend
on it. Figure~\ref{w_c_p}(a) indicates that, in the region of
interest, the average decay time from the low to the high CO
coverage phase, $\langle\tau_\text{p}\rangle$, while being
dependent on the final pressure $y$, is fairly independent of the
pressure at which the system is prepared, $y_\text{w}$.
In the following we then take $y_\text{w}=0.45$ to calculate
$\tau_\text{p}$. It is also necessary to evaluate the dependence
of $\langle\tau_\text{p}\rangle$ on the selected 
cut-off value $\theta_\text{CO}^*$. In Fig.~\ref{w_c_p}(b) we
plot $\langle\tau_\text{p}\rangle$ vs 
$\theta_\text{CO}^*$ for different values of $k$ and $y$. Clearly,
$\langle\tau_\text{p}\rangle$ increases with $\theta_{\rm CO}^*$,
however there is a region where it is relatively weakly dependent
on the cutoff. We choose to perform our measurements of $\langle
\tau_\text{p}\rangle$ at $\theta_{\rm CO}^{*}=0.65$, well inside this
region. 

\begin{figure}
\centering\includegraphics[scale=.5, clip]{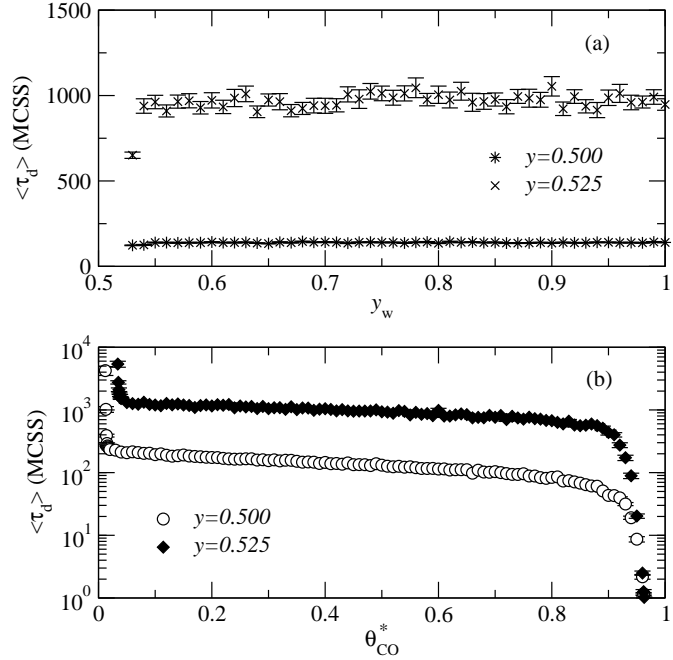}
\caption[]{$\langle\tau_\text{d}\rangle$ as a function of (a) $y_\text{w}$ with
$\theta_\text{CO}^*=0.45$, and (b) $\theta_\text{CO}^*$
with $y_\text{w}=0.57$; for $k=0.02$, $L=100$ and two different values of $y$.}
\label{w_c_d}
\end{figure}

Figure~\ref{w_c_d} indicates that the average decay times
associated with the decontamination of the CO
surface, $\langle \tau_\text{d} \rangle$, i.e.\ 
the relaxation time from the high CO
coverage phase  to the low-coverage phase, behaves in a similar way
to $\langle \tau_\text{p} \rangle$. 
Figure~\ref{w_c_d}(a) clearly indicates that,
to an even higher degree than in the poisoning process,
$\langle \tau_\text{d} \rangle$ is independent of $y_\text{w}$. We choose
$y_\text{w}=0.57$ for our calculations. Figure~\ref{w_c_d}(b)
indicates that, as expected, $\langle\tau_\text{d}\rangle$
increases as $\theta_{\rm CO}^*$ decreases, but there is a
range of values of the cut-off
 where the dependence is relatively small. In the following we
calculate $\langle\tau_\text{d}\rangle$ with
$\theta_{\rm CO}^{*}=0.45$, which lies in this region.

\begin{figure}
\centering\includegraphics[scale=.7]{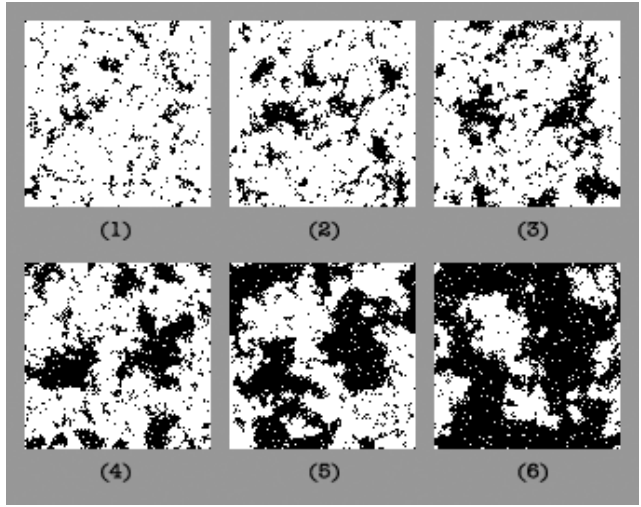}
\caption[]{Snapshot configurations obtained at different 
(unevenly spaced) times, after a
sudden change of $y$ from $y_\text{w}=0.45$
to $y = 0.538 > y_{2}(k)$, i.e., $\Delta^{-1}\approx 200$. 
For $k=0.02$ and $L=100$.
Here and in Figs.~\ref{selected_d_MD}, \ref{selected_p_SD}, 
and \ref{selected_d_SD},
the black dots represent lattice sites occupied by CO,
while both adsorbed O atoms and empty lattice sites are white.}
\label{selected_p_MD}
\end{figure}

\begin{figure}
\centering\includegraphics[scale=.7]{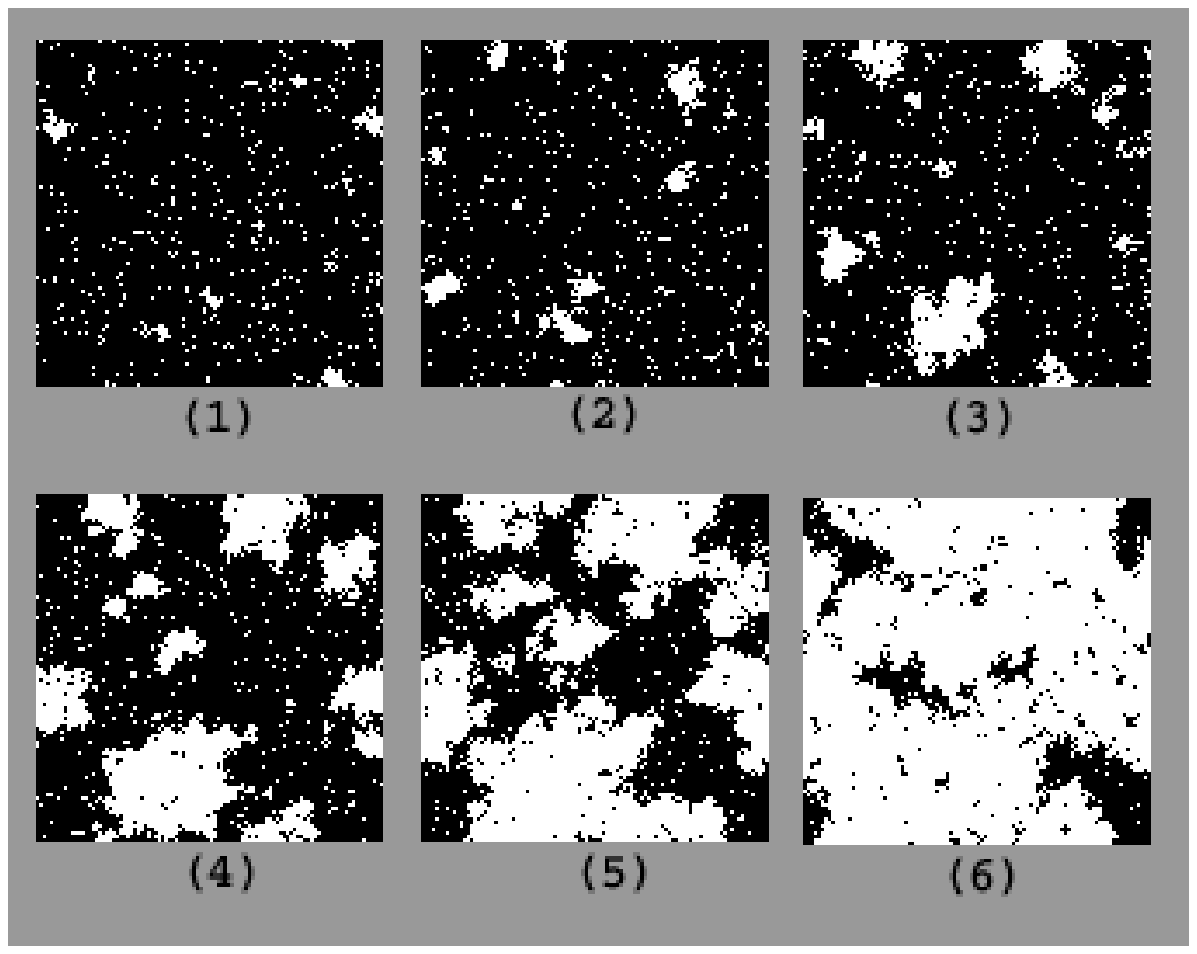}
\caption[]{Snapshot configurations obtained at different 
(unevenly spaced) times, after
a sudden change of $y$ from $y=0.65=y_\text{w}$
to $y = 0.5232 < y_{2}(k)$, i.e., $\Delta^{-1}\approx 100$. 
For $k=0.02$ and $L=100$.}
\label{selected_d_MD}
\end{figure}

\begin{figure}
\centering\includegraphics[scale=.7]{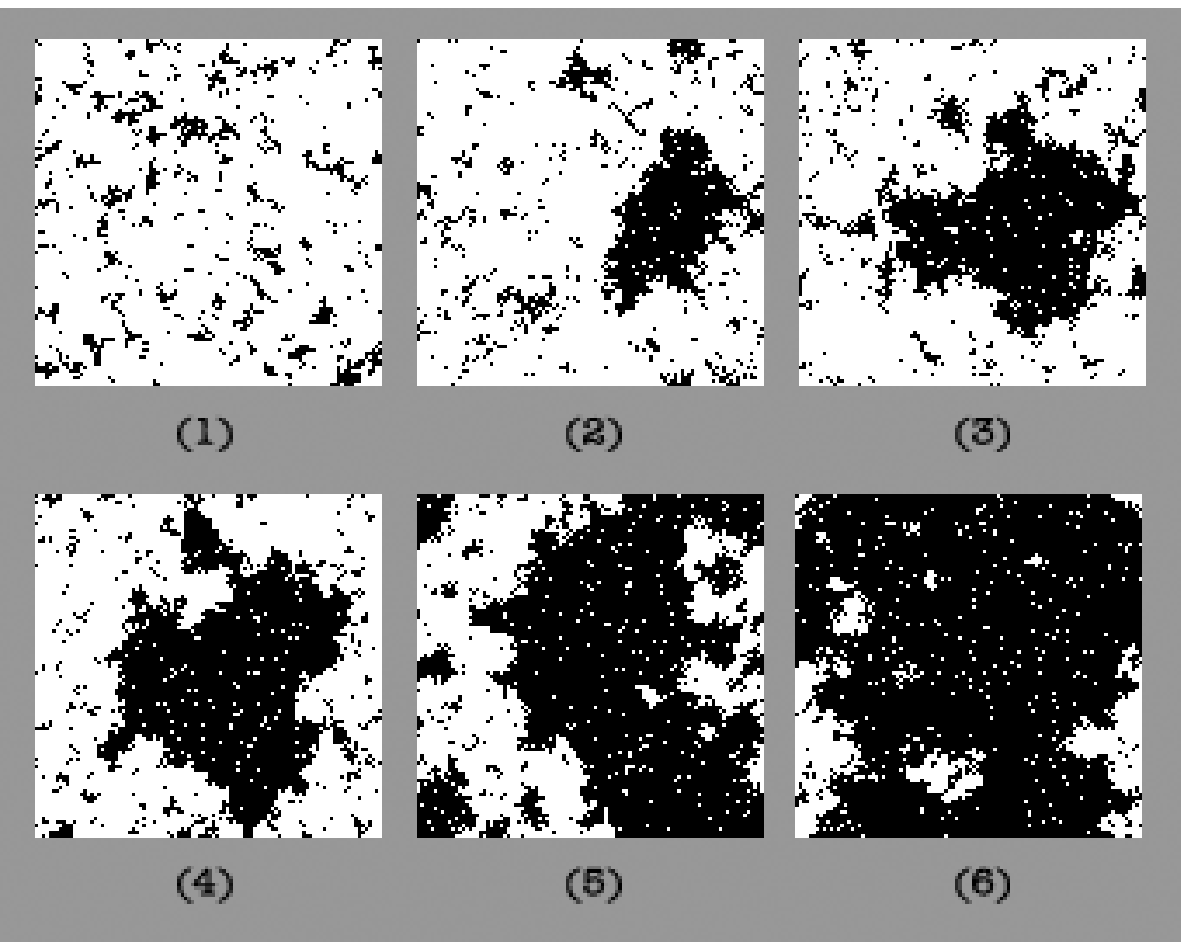}
\caption[]{Snapshot configurations obtained at different 
(unevenly spaced) times, after
a sudden change of $y$ from $y=0.45=y_\text{w}$
to $y = 0.5338 > y_{2}(k)$, i.e., $\Delta^{-1}\approx 2000$. 
For $k=0.02$ and $L=100$.}
\label{selected_p_SD}
\end{figure}

\begin{figure}
\centering\includegraphics[scale=.7]{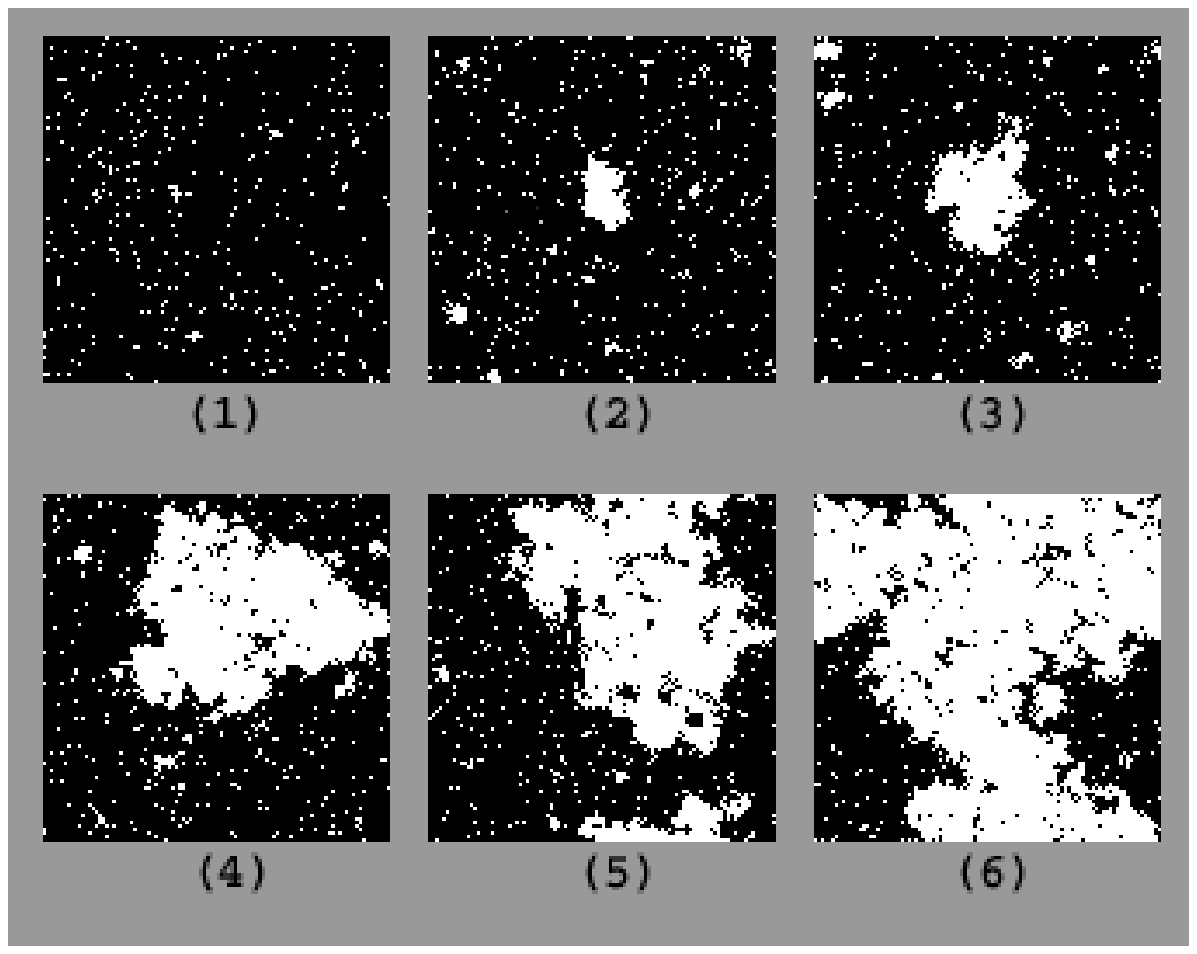}
\caption[]{Snapshot configurations obtained at different 
(unevenly spaced) times, after
a sudden change of $y$ from $y=0.65=y_\text{w}$
to $y = 0.5315 < y_{2}(k)$, i.e., $\Delta^{-1}\approx 500$. 
For $k=0.02$ and $L=100$.}
\label{selected_d_SD}
\end{figure}

\begin{figure}
\centering\includegraphics[scale=.5, clip]{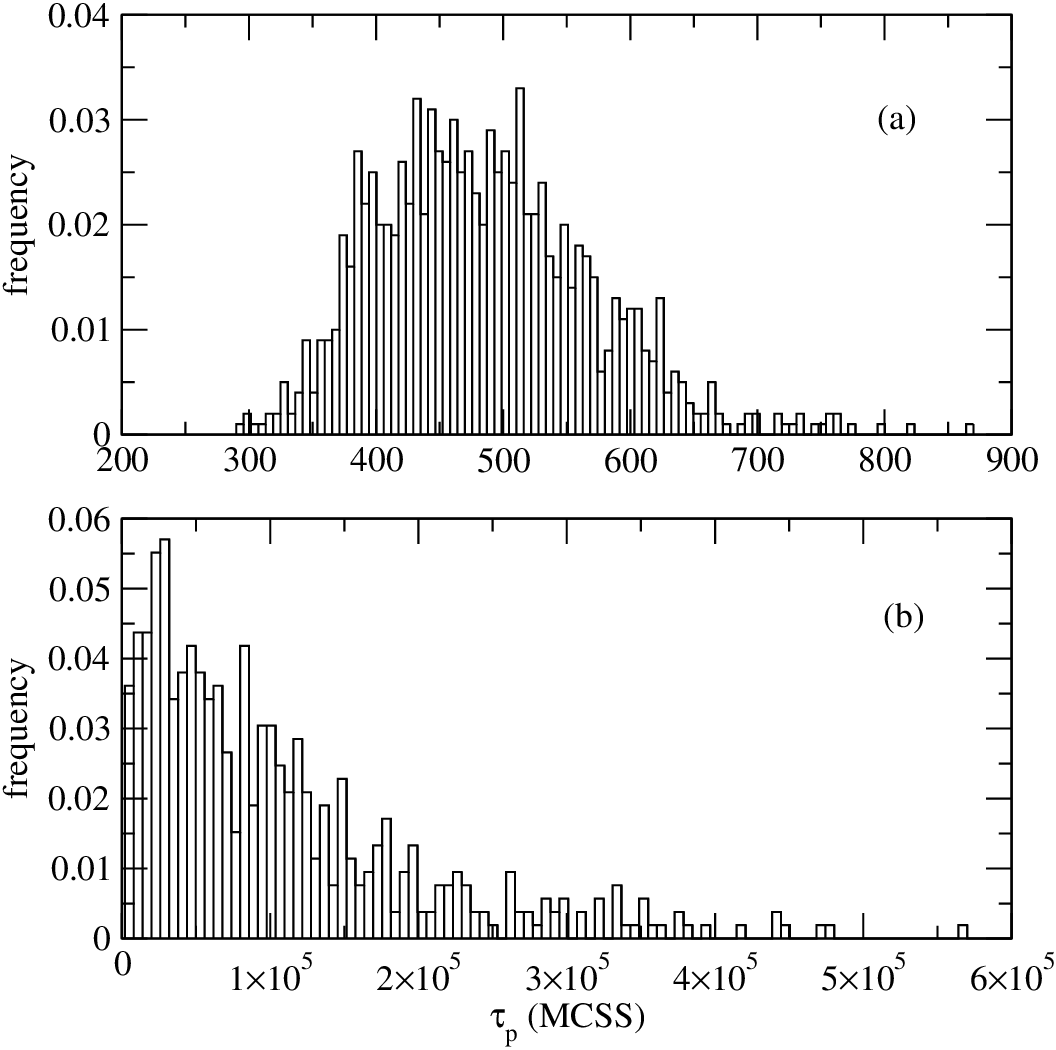}
\caption{Plot of distributions of $\tau_\text{p}$ for $k=0.02$ and $L=100$,
(a) $\Delta^{-1}\approx 200$ and (b) $\Delta^{-1}\approx 2000$.
Note the very different time scales in (a) and (b).}
\label{histo_p}
\end{figure}

\begin{figure}
\centering\includegraphics[scale=.5, clip]{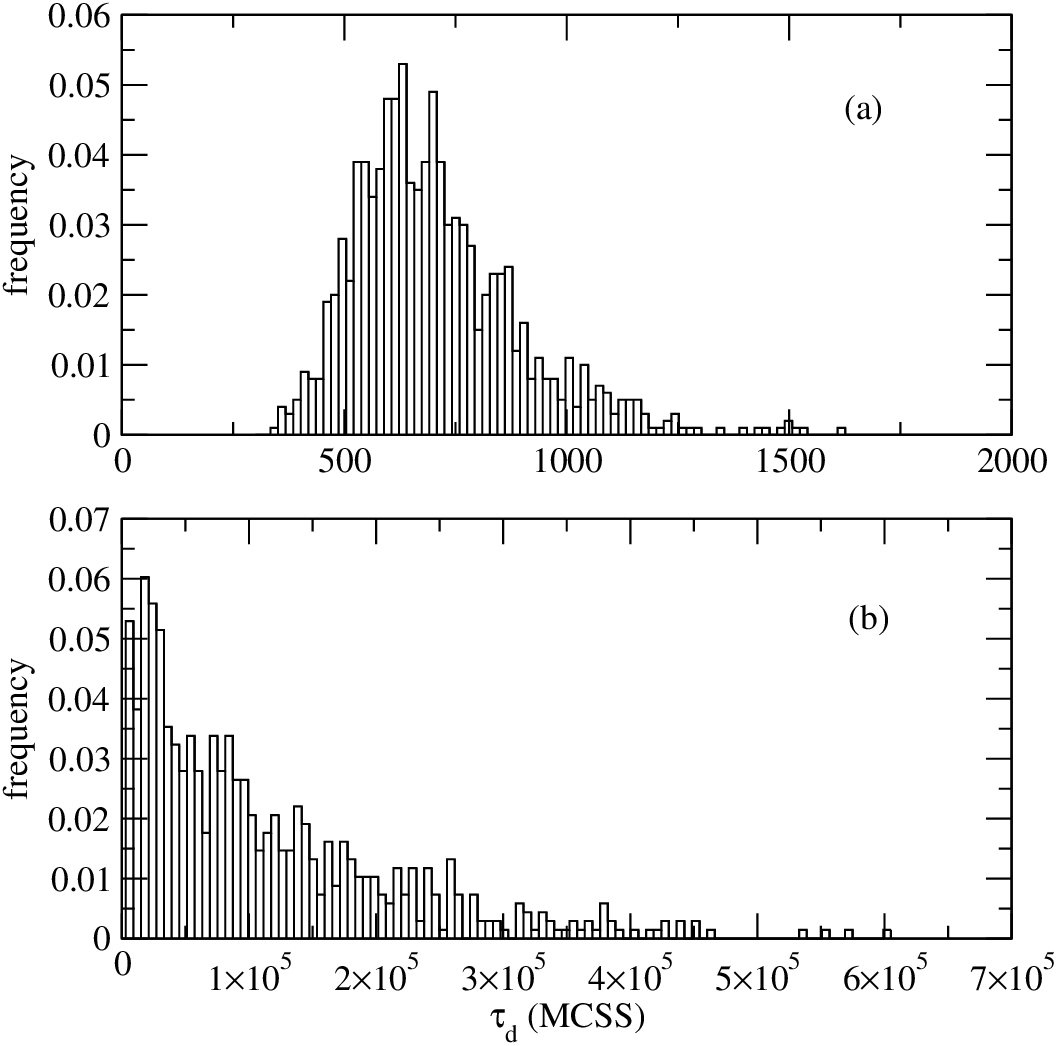}
\caption{Plot of distributions of $\tau_\text{d}$ for $k=0.02$ and $L=100$,
(a) $\Delta^{-1}\approx 100$ and (b) $\Delta^{-1}\approx 500$.
Note the very different time scales in (a) and (b).}
\label{histo_d}
\end{figure}

We define the quantity
\begin{equation}
\Delta=|y-y_{2}| \;, 
\end{equation}
which measures how far the system is
from the coexistence curve and depends on $k$ and $L$ through $y_2$.
In Figs.~\ref{selected_p_MD} and \ref{selected_d_MD} we present snapshot
configurations obtained during the relaxation from the low to
the high coverage phase and from the high to the low coverage
phase, respectively, when $\Delta^{-1}$ is small, i.e., far from the transition.
In Figs.~\ref{selected_p_SD} and  \ref{selected_d_SD}
we show snapshot configurations for a much larger value of 
$\Delta^{-1}$, i.e., close to the transition.  The difference in the decay
mechanisms is evident from the figures. Figures~\ref{selected_p_MD}
and \ref{selected_d_MD} clearly suggest
that when $\Delta^{-1}$ is small, the system decays by
nucleation and growth of multiple droplets of the stable phase. 
In contrast, Figs.~\ref{selected_p_SD} and  \ref{selected_d_SD} show that when
$\Delta^{-1}$ is large, the system decays by nucleation and growth of a
single droplet of the stable phase, which eventually takes over the
entire system. The probability distributions of $\tau_\text{p}$ and
$\tau_\text{d}$, shown in Figs.~\ref{histo_p} 
and \ref{histo_d}, respectively, also indicate
clear differences between the statistics of the decay times near and
far from the coexistence line. Far from coexistence ($\Delta^{-1}$ small),
the decay times follow an approximately Gaussian distribution. 
In contrast, near the
transition ($\Delta^{-1}$ large), the distribution is approximately exponential. 


The statistics of the metastable lifetimes in the model studied 
here are strikingly similar to those found in Hamiltonian systems 
that decay toward thermodynamic equilibrium from a metastable 
phase associated with an {\it equilibrium\/} 
phase transition. Well-studied examples are metastable 
decay in kinetic Ising \cite{RIKV94,RICH95,RAMO99,KOLE03} 
and lattice-gas models \cite{NOVO00,FRAN04} with such applications 
as magnetism switching and submonolayer adsorption. 
In the present paper we will for simplicity refer to this 
latter case as the ``Hamiltonian" case, thus emphasizing the 
lack of a Hamiltonian and the consequent lack of a concept of thermodynamic 
equilibrium for the system studied in the present paper. 
The metastable decay in such a Hamiltonian system occurs via 
nucleation and subsequent
growth of ``droplets" inside which the order parameter is 
close to its equilibrium value,
and it is well described by the classic 
KJMA theory of
phase transformation \cite{KOLM37,JOHN39,AVRAMI}. 
The basic assumption of this theory is that droplets of the stable 
phase nucleate in a Poisson process
at a rate $I$ per unit volume. After nucleation, the droplet 
radius is assumed to grow with a constant speed, $v$.
If the first droplet to nucleate grows fast enough to fill the system  
before another is likely to nucleate, then it completes the 
phase transformation by itself -- a process known as single-droplet (SD) decay. 
However, if the growth is slow, 
so that many droplets can nucleate within the time it would take a 
single droplet to fill the system, the phase transformation will 
proceed via a large number of droplets that nucleate and grow in 
parallel -- a process known as multidroplet (MD) decay. 
This simple observation 
can be turned into a formal scaling argument by constructing 
the characteristic length $R_0 = (v/I)^{1/3}$, which is a 
measure of the average size to which a droplet will grow before it 
touches another droplet. 
(Results are here reviewed only for $d=2$. Results for general $d$
can be found in, e.g., Refs.~\cite{RIKV94,RAMO99} and references therein.)
If $R_0 \gg L$, the system is in the SD regime, and the 
metastable lifetime is simply the average time between 
nucleation events, $\tau_{\rm SD} \approx 1/(I L^2)$. 
Since the nucleation events constitute a Poisson process, 
$\tau_{\rm SD}$ follows an exponential distribution, 
similar to the ones shown in Figs.~\ref{histo_p}(b) and~\ref{histo_d}(b) 
for the system discussed here. 
If, on the other hand, $R_0 \ll L$, then the system is in the 
MD regime, and the metastable lifetime is obtained as 
$\tau_{\rm MD} = R_0/v = 1/(v^2 I)^{1/3}$. 
Since the metastable decay in this case consists of a large 
number of droplets that nucleate and grow independently, 
$\tau_{\rm MD}$ follows a Gaussian distribution with a standard 
deviation proportional to $R_0/L$, 
similar to the ones shown in Figs.~\ref{histo_p}(a) and~\ref{histo_d}(a) 
for the system discussed here. 

The SD regime with its exponential lifetime distribution is a subregion
of a broader {\it stochastic regime\/}, while the MD regime with its narrow 
Gaussian distribution is part of a broader {\it deterministic regime\/}. 
The relative standard deviation of the lifetimes is defined by
\begin{equation}
r=\frac{\sqrt{\langle\tau^2\rangle-\langle\tau\rangle^2}}{\langle\tau\rangle} 
\approx 
\left\{
\begin{array}{ll}
R_0/L < 1 
& \mbox{{\rm in MD regime}} \\
1               
& \mbox{{\rm in SD regime}}
\end{array}
\right.
\;.
\label{rel_dev_std}
\end{equation}
The limit between the SD and MD regimes is called the Dynamic Spinodal 
(DSP) and corresponds to $R_0 \approx L$. It is, however, easier to estimate 
it as given by the values of $I$ and $v$ that yield $r = 1/2$ \cite{RIKV94}. 

So far, the KJMA results discussed do not require a specific dependence
of the nucleation rate $I$ and growth velocity $v$ on the macroscopic
control parameters, which for metastable decay in Hamiltonian systems
are the applied magnetic field $H$ (or chemical potential or supersaturation 
for lattice-gas
models) and the temperature $T$. In the present model the analogous
quantities should be the distance from coexistence $\Delta$ and the
desorption rate $k$. Hamiltonian systems are described by a free
energy, and standard arguments of droplet theory show that for not too
strong fields, 
$I \sim \exp \left[ - c(T)/(T |H|)\right]$, where $c(T)$ is
well approximated as proportional to the equilibrium interface tension
between the metastable and equilibrium phases. For weak fields, $v
\propto |H|$ -- an effect that to a reasonable approximation can be
ignored compared to the exponential dependence on $1/|H|$ in $I$. 

The present nonequilibrium system has no Hamiltonian and so no
free-energy function. However, let us for the moment postulate that 
$I(\Delta,k) \sim \exp(-c(k)/\Delta)$ for reasonably small $\Delta$, 
and that $v(\Delta,k)$ depends
comparatively weakly on its parameters so that it can be taken as
approximately constant. Following the method of data analysis introduced
in Ref.~\cite{RIKV94}, we then expect logarithmic plots of $\langle \tau_{\rm
p} \rangle$ and $\langle \tau_{\rm d} \rangle$ (and of $r$ in the MD regime) 
vs $1/\Delta$ to be approximately linear. Furthermore, the general
KJMA arguments given above indicate that the lifetimes should be
independent of $L$ in the MD regime and $\propto L^{-2}$ in the SD
regime.   

\begin{figure}
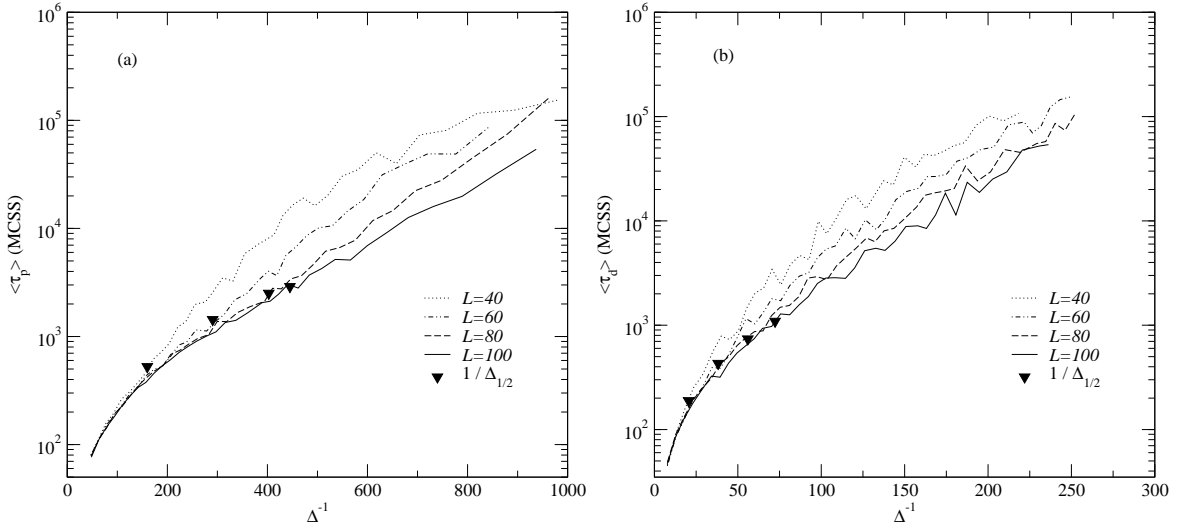

\centering\includegraphics[scale=.4, clip]{fig17a.eps}
\centering\includegraphics[scale=.4, clip]{fig17b.eps}
\caption[]{Log-linear plot of  
(a) $\langle\tau_\text{p}\rangle$ vs $\Delta^{-1}$
and (b) $\langle\tau_\text{d}\rangle$ vs $\Delta^{-1}$,
shown for $k=0.01$ and different values of $L$.
The solid inverted triangles indicate $\Delta_{1/2}^{-1}$ for each value
of $L$. Approximate data collapse is seen for 
$\Delta^{-1} < \Delta_{1/2}^{-1}$. 
} 
\label{log_tau_vs_inv_dy1}
\end{figure}

\begin{figure}
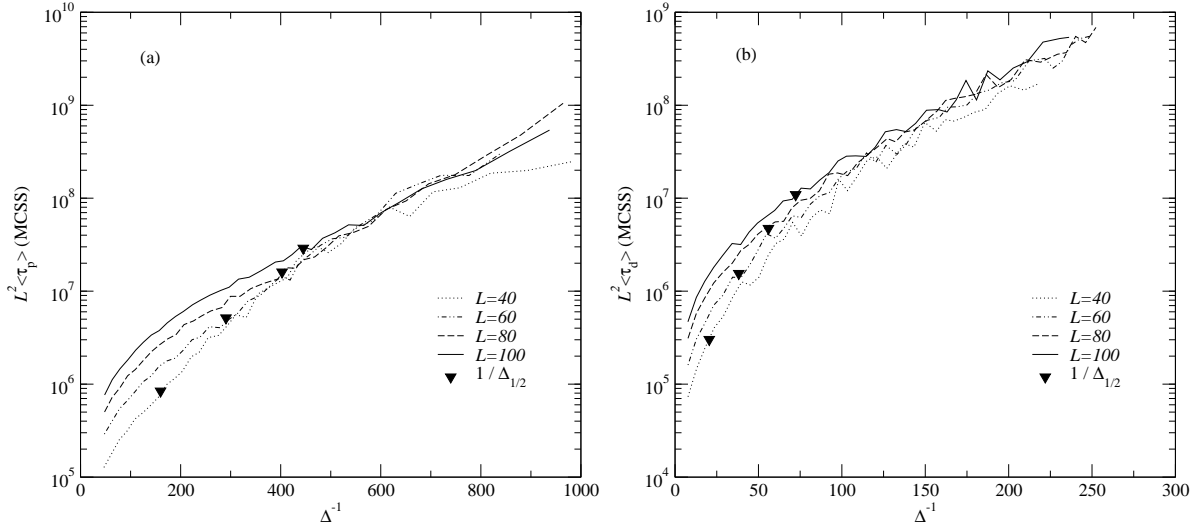

\centering\includegraphics[scale=.4, clip]{fig18a.eps}
\centering\includegraphics[scale=.4, clip]{fig18b.eps}
\caption[]{Log-linear plot of (a) $L^2\langle\tau_\text{p}\rangle$ vs $\Delta^{-1}$
and (b) $L^2\langle\tau_\text{d}\rangle$ vs $\Delta^{-1}$,
shown for $k=0.01$ and different values of $L$.
The solid inverted triangles indicate $\Delta_{1/2}^{-1}$ for each value
of $L$. Approximate data collapse is seen for 
$\Delta^{-1} > \Delta_{1/2}^{-1}$. 
} 
\label{FSS_tau1}
\end{figure}

Figure~\ref{log_tau_vs_inv_dy1} 
shows log-linear plots of $\langle\tau_\text{p}\rangle$ 
and $\langle\tau_\text{d}\rangle$
vs $\Delta^{-1}$ for $k=0.01$ and different sizes. 
Similar plots were also obtained for $k=0.02$ (not shown). 
The plots clearly indicate that there is a regime,
corresponding to small $\Delta^{-1}$, where $\langle\tau_\text{p}\rangle$ is
independent of $L$. 
Figure~\ref{FSS_tau1}
strongly indicates that
when $\Delta^{-1}$ is large, the decay times are inversely
proportional to $1/L^2$. Only the data for $L=40$ do not seem
to follow this dependence. We believe it is possible that $L=40$ 
becomes smaller than the critical droplet size for large $\Delta^{-1}$
(incipient ``coexistence regime," see Ref.~\cite{RIKV94}).

\begin{figure}
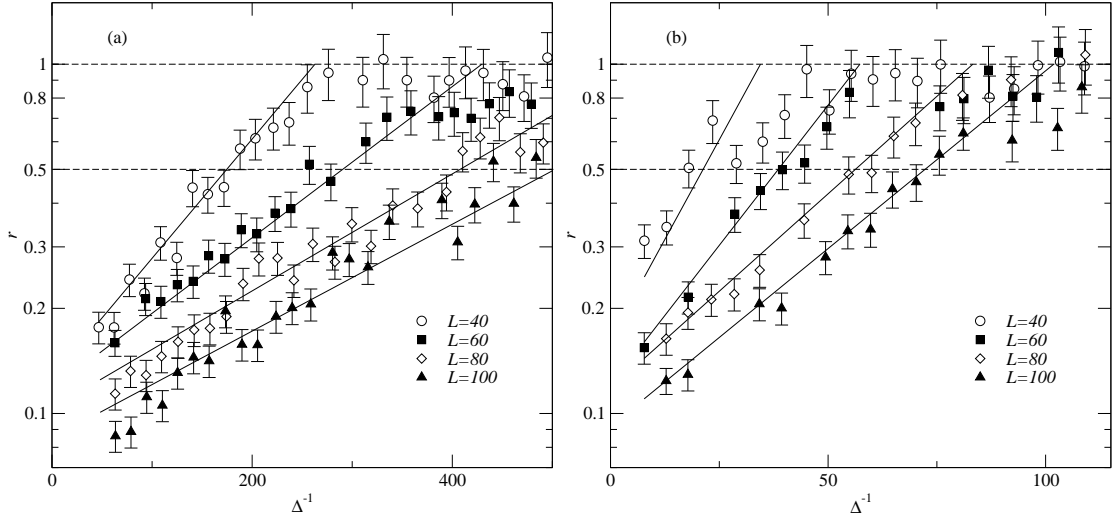

\centering\includegraphics[scale=.4, clip]{fig19a.eps}
\centering\includegraphics[scale=.4, clip]{fig19b.eps}
\caption[]{The relative standard deviation $r$ of 
(a) $\langle \tau_\text{p}\rangle$ and (b) $\langle \tau_\text{d}\rangle$, shown on a
logarithmic scale vs $\Delta^{-1}$ for $k=0.01$. The behavior of $r$
crosses over from the approximate straight line expected from our
postulate in the deterministic regime to $r\approx 1$ in the 
stochastic regime. The solid lines
are weighted least-squares fits to the data in the linear region.}
\label{r_vs_inv_dy1}
\end{figure}

\begin{figure}
\centering\includegraphics[scale=.5, clip]{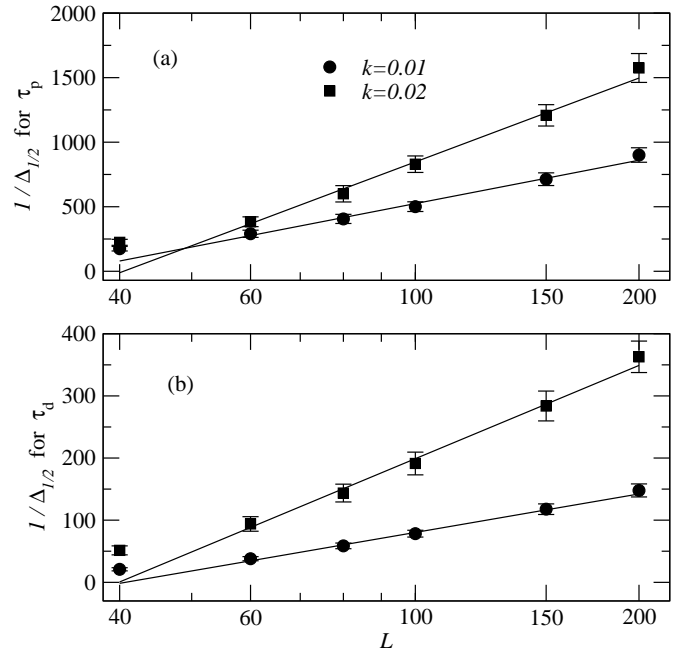}
\caption[]{The estimates $1/\Delta_{1/2}$ for $1/\Delta_\text{DSP}$ 
for (a) poisoning and (b) decontamination as obtained
from Fig.~\ref{r_vs_inv_dy1}
and analogous data, shown vs $L$ on a logarithmic scale. 
The solid lines are weighted least-squares fits excluding
the points corresponding to $L=40$.}
\label{dsp}
\end{figure}

To further explore the applicability of the KJMA theory 
and our postulate to the model, we next 
determine the dynamic spinodal, $\Delta_\text{DSP}$, 
that separates the stochastic and the deterministic regimes.  We 
calculate the relative standard deviation $r$ of Eq.~(\ref{rel_dev_std}), 
which is shown on a logarithmic scale 
vs $\Delta^{-1}$
in Fig.~\ref{r_vs_inv_dy1},
where error bars are estimated by standard error-propagation methods as
\begin{equation}
\sigma_r \approx 
\frac{r}{\sqrt{n-1}}\left(1+\frac{n-1}{n}r^2\right)^{\frac{1}{2}}.
\end{equation}
As can be seen, $r$ crosses over from the approximately linear behavior 
expected from our postulate 
for small $\Delta^{-1}$ to $r\approx 1$ for larger $\Delta^{-1}$.
We take as our
estimate for $\Delta_\text{DSP}$ the value $\Delta_{1/2}$ for which $r=0.5$.
This crossover 
is determined for each value of $L$ from the crossing of a weighted
least-squares fit to $\ln r$ in the linear region of
Fig.~\ref{r_vs_inv_dy1}
with the horizontal line
$r=0.5$. The resulting estimates are shown in
Fig.~\ref{dsp} as $1/\Delta_{1/2}$ vs
$L$ on a logarithmic scale, with error bars estimated from those in
Fig.~\ref{r_vs_inv_dy1}
by standard error-propagation methods.
The numerical results are consistent with the analytical prediction  
based on our postulate,
\begin{equation}
\Delta_\text{DSP}\sim\ 1/ \ln L .
\end{equation}

\begin{figure}
\centering\includegraphics[scale=.5, clip]{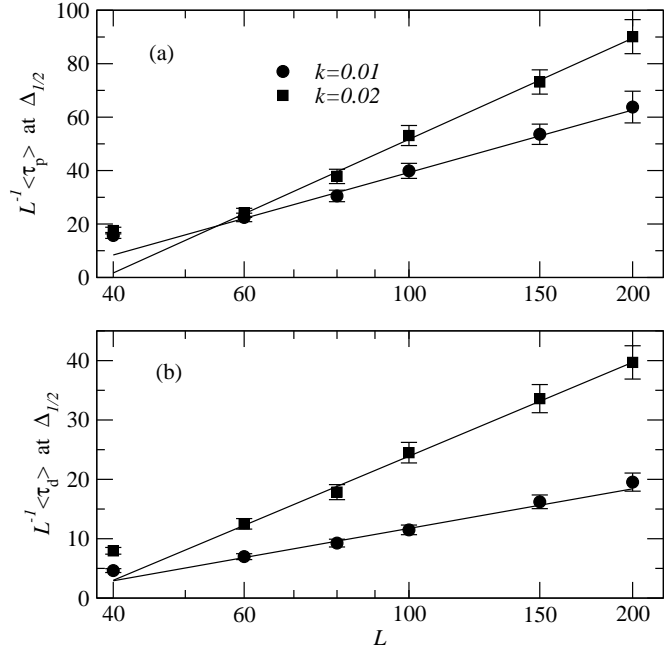}
\caption[]{(a) $\langle \tau_\text{p}\rangle/L$ at $1/\Delta_{1/2}$
and (b) $\langle \tau_\text{d}\rangle/L$ at $1/\Delta_{1/2}$,
shown vs $L$ on a logarithmic scale. The solid straight lines
are weighted least-squares fits. Excluding the points corresponding to
$L=40$, the figures supports the asymptotic behavior
$\langle \tau_\text{p}\rangle \propto L\ln L$ at the DSP.}
\label{tau_at_dsp}
\end{figure}

The asymptotic $L$ dependence of the lifetime at the DSP, analogously given by
\begin{equation} 
\langle \tau \rangle 
\propto L/\Delta_\text{DSP}\sim L\left( \ln L\right),
\end{equation}
is illustrated in Fig.~\ref{tau_at_dsp}.
For each $L$ this lifetime was obtained by interpolation between
the two closest field values bracketing $\Delta_{1/2}$, 
for which simulations had
been performed. The uncertainty in the resulting estimate was obtained by
standard error propagation, taking the standard deviation in
$\langle \tau \rangle$ from  Eq.~(\ref{rel_dev_std})
with $r=1/2$, and $\langle \tau \rangle $ from
Fig.~\ref{log_tau_vs_inv_dy1}.

While a direct analogue of the surface tension does
not exist in the present system, the results described above 
strongly suggest that it obeys a decay mechanism very similar
to the one described by the standard KJMA theory of phase transformation
by nucleation and growth, which predicts well-defined single-droplet and 
multidroplet regimes.
A significant difference between our Fig.~\ref{log_tau_vs_inv_dy1} 
and analogous figures showing the
metastable lifetime for a Hamiltonian system vs inverse field or 
supersaturation (see, e.g, Fig.~2 of Ref.~\cite{RIKV94} and Fig.~2 of 
Ref.~\cite{KOLE03}), is that we here see no marked change in the slope of
the curves at the DPT. One possible explanation is that the ``effective
surface tension'' in the present case may decrease substantially with
increasing $\Delta$, in contrast to the situation in Hamiltonian
systems. 

\begin{figure}
\centering\includegraphics[scale=.5, clip]{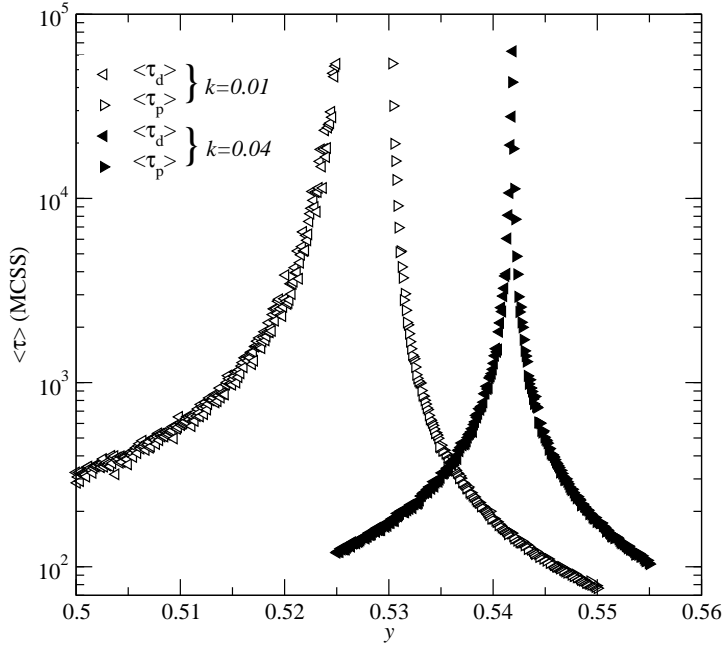}
\caption[]{Decay times
as functions of $y$ when the system evolves toward the low CO
coverage region ($\theta_\text{CO}^*=0.45, y_\text{w}=0.55$, 
left-pointing triangles) and
when it evolves toward the high CO coverage region
($\theta_\text{CO}^*=0.65, y_\text{w}=0.475$, 
right-pointing triangles) for $k=0.01$ and $k=0.04$, and $L=100$.
As discussed in the text, the divergence is exponential in 
$1/|y - y_2(k,L)| = 1/\Delta$. 
}
\label{decay}
\end{figure}

The decay times increase as $y$ approaches the transition line
$y_{2}(k)$, however their behaviors depend on the direction of
approach to the transition value, as can be seen in Fig.~\ref{decay}. 
In a previous work we have shown how this asymmetry can be exploited to 
enhance the catalytic activity by subjecting the system to periodic 
variation of the external pressure with periods related to the decay 
times in each direction \cite{machado04}.

The asymmetry between the decay times when the system evolves to
the low CO coverage phase, $\langle \tau_\text{d} \rangle$, and the decay time
toward the high CO coverage phase, $\langle \tau_\text{p} \rangle$, 
becomes more
evident as the desorption parameter $k$ decreases, as can also be
seen in Fig.~\ref{decay}. The extreme case occurs  when $k=0$,
where $\langle \tau_\text{d} \rangle = \infty$, independently of the
value of $y$, since in the high CO coverage
phase the surface then becomes irreversibly poisoned with CO, whereas
$\langle \tau_\text{p} \rangle$ remains finite and dependent on $y$.

In Fig.~\ref{decay} it can be seen that $\langle \tau_\text{d}
\rangle$ and
$\langle \tau_\text{p} \rangle$ appear to diverge at the same value of $y$.
At this point the system spends, on average, the same amount of
time in the low and high CO coverage phases, and it agrees with the
value for the coexistence point $y_2$, calculated from the
order-parameter distribution in Sec.~\ref{sec:CC}. Since at this point
the average decay time is of the order of $10^4$ MCSS, during the
observation time of $10^6$ MCSS it is very probable that several
transitions occur between the two phases. These transitions are
observed in Fig.~\ref{histos3}, where it can be seen that the probability
distribution, $P(\theta_{CO})$ is bimodal and quite symmetric,
indicating that $\langle \tau_\text{d} \rangle \approx 
\langle \tau_\text{p} \rangle$ along the coexistence curve for finite $L$.

\section{Conclusions}
\label{sec:conc}

We have investigated by kinetic Monte Carlo simulation
the dynamical behavior of a ZGB model with
desorption near the coexistence curve between the active and the CO
poisoned nonequilibrium phases.  
We perform an extensive finite-size scaling analysis of 
the fluctuations and of the fourth-order reduced cumulant of the CO
coverage, which plays the role of an order parameter.
Our results strongly indicate, as also previously
suggested by others, that the system undergoes a first-order nonequilibrium 
phase transition between
the active and the CO poisoned phases. The coexistence
curve terminates at a critical value of the desorption rate.
We also calculated several points on the coexistence curve.

Next we calculated the system-size dependence of the decay times of
the metastable phases when the system is driven into the CO poisoned
phase from the active phase, and {\it vice versa\/}. 
We found that near the coexistence curve
the decay times are inversely proportional to $1/L^2$, and the decay
mechanism consists of the nucleation and growth of a single supercritical 
droplet of the stable
phase. In contrast, far from the coexistence curve, the decay times are
independent of the system size, and the decay proceeds by random nucleation
of many droplets of the stable phase that grow independently 
and coalesce. These
regimes are separated by a dynamic spinodal that vanishes logarithmically
with system size. These results strongly suggest that our
nonequilibrium, non-Hamiltonian 
system follows a decay mechanism very similar to the one described by the 
classic KJMA theory of phase transformation by nucleation and growth
near a first-order equilibrium phase transition, 
which predicts well-defined single-droplet and multidroplet regimes. 
In the present far-from-equilibrium system,
the desorption parameter and the distance to the
coexistence point play the roles of the temperature and the external field
or supersaturation, respectively.
Very recently, indications of KJMA behavior have also been
observed in another non-Hamiltonian, nonequilibrium system: an
ecological model of invasion by exotic species \cite{KORN05}. 
We find quite exciting the strong similarity between the dynamics of 
metastable decay in far-from-equilibrium, non-Hamiltonian systems of 
applied importance, and
the well-known behavior in systems that can be described by a Hamiltonian.

\section*{Acknowledgments}

We are grateful to R.~M.\ Ziff for insightful comments. 
This work was
supported in part by U.S.\ National Science Foundation Grant No.\ DMR-0240078,
by Florida State University through the Center from Materials Research and
Technology, the School of Computational Science, and the National High Magnetic
Field Laboratory, and by the Deanship of Research and Development
of Universidad Sim\'on Bol\y var.

\end{document}